\documentclass[pra,twocolumn,floatfix,aps,psfig]{revtex4}
\usepackage{amssymb}
\usepackage{graphicx}
\usepackage{subfigure}

\newcommand{\beq}{\begin{equation}}
\newcommand{\eeq}{\end{equation}}
\newcommand{\beqa}{\begin{eqnarray}}
\newcommand{\eeqa}{\end{eqnarray}}
\newcommand{\ba}{\begin{array}}
\newcommand{\ea}{\end{array}}

\begin{document}

\title{Application of the Feshbach-resonance management \\
to a tightly confined Bose-Einstein condensate}
\author{G. Filatrella,$^{1,2}$ B.A. Malomed$^{3}$ and L. Salasnich$^{4}$}
\affiliation{$^1$Department of Biological and Environmental Sciences, University of
Sannio, via Port'Arsa 11, 82100 Benevento, Italy \\
$^2$CNR-INFM, Regional Laboratory SUPERMAT, via S. Allende, 84081 Baronissi,
Italy \\
$^3$Department of Physical Electronics, School of Electrical Engineering,
Faculty of Engineering, Tel Aviv University, Tel Aviv 69978, Israel \\
$^4$CNR-INFM, CNISM, and Department of Physics ``Galileo Galilei'',
University of Padua, 35122 Padua, Italy}

\begin{abstract}
We study suppression of the collapse and stabilization of matter-wave
solitons by means of time-periodic modulation of the effective nonlinearity,
using the nonpolynomial Schr\"{o}dinger equation (NPSE) for BEC trapped in a
tight cigar-shaped potential. By means of systematic simulations, a
stability region is identified in the plane of the modulation amplitude and
frequency. In the low-frequency regime, solitons feature chaotic evolution,
although they remain robust objects.
\end{abstract}

\maketitle

\section{Introduction}

Dilute atomic Bose-Einstein condensates (BECs) are accurately described by
the Gross-Pitaevskii equation (GPE), alias the cubic nonlinear Schr\"{o}%
dinger equation (NLSE) \cite{leggett}. The sign of the cubic term in the GPE
corresponds to the self-defocusing or focusing, if interactions between
atoms in the condensate are characterized, respectively, by the positive or
negative \textit{s}-wave scattering length. The self-focusing GPE in any
dimension (1D, 2D, or 3D) gives rise to soliton solutions, which are stable
in the 1D case. The creation of 1D matter-wave solitons has been reported in
experimental works \cite{soliton}, while 2D and 3D solitons are unstable
against the critical and supercritical collapse, respectively (these 2D
states are usually called \textit{Townes solitons}, TSs) \cite{Berge'}. It
was predicted that TSs may be stabilized in the framework of the 2D GPE,
without using an external potential, if the constant scattering length is
replaced by a time-dependent one, that periodically changes its sign \cite%
{book-boris}. In BEC, this can be implemented by means of the \textit{%
Feshbach-resonance management} (FRM), i.e., by applying a low-frequency ac
magnetic field which acts via the Feshbach resonance \cite{Feshbach}. This
stabilization mechanism was demonstrated in optics, in terms of the
transmission of a light beam through a bulk medium composed of layers with
alternating signs of the Kerr nonlinearity \cite{Isaac}, and then in the
framework of the 2D GPE \cite{2Dstabilization,Spain}. A somewhat similar
technique was proposed recently, making use of a linear coupling, induced by
means of a resonant electromagnetic wave, between two different hyperfine
states of atoms, which feature opposite signs of the scattering length \cite%
{Randy}. The analysis of the FRM was extended to include averaging
techniques \cite{averaging}, generation of solitons from periodic waves \cite%
{periodic}, the stabilization of higher-order solitons \cite{higher-order},
management of discrete arrays \cite{discrete}, and the case of a chirped
modulation frequency \cite{Nicolin}. However, the stabilization based on the
FRM may be, strictly speaking, a transient dynamical regime, as extremely
long simulations suggest that the FRM-stabilized TS may be subject to a very
slow decay \cite{Japan}.

The stabilization of 3D solitons by means of the FRM technique alone
is not possible, but stable 3D solitons were predicted in a model
combining the FRM and a 1D periodic potential \cite{Warsaw}.
Similarly, the stabilization is possible when the FRM is applied in
combination with a parabolic potential which strongly confines the
condensate in one direction \cite{Spain}. Most relevant to the
experiment is the ``cigar-shaped" setting, with the BEC tightly
confined in two transverse directions, while the third direction
remains free \cite{soliton}. In the usual approximation, with the
cubic nonlinearity in the corresponding 1D GPE, the analysis of the
FRM in the latter setting amounts to that reported in Refs.
\cite{Feshbach}. However, if the density of the condensate is not
very low, the description in terms of the cubic nonlinearity is
inappropriate, the respective 1D equation taking the form of the
\textit{nonpolynomial Schr\"{o}dinger equation} (NPSE). In
particular, it admits the onset of the collapse in the
self-attractive condensate in the framework of the 1D description \cite%
{sala1}. Accordingly, a relevant problem, which is the subject of the
present work, is to study the possibility of the collapse suppression by
means of the FRM technique in the framework of the 1D NPSE. It is relevant
to mention that the NPSE was recently used to describe Faraday waves
generated in the cigar-shaped trap by a time-periodic modulation of the
strength of the transverse confinement \cite{Ricardo}. We introduce the
model in Section 2, and report results obtained by means of systematic
numerical simulations in Section 3.

\section{The nonpolynomial Schr\"{o}dinger equation}

The normalized form of the 3D GPE with the transverse harmonic trapping
potential, which acts in the $(x,y)$ plane, is
\begin{eqnarray}
i\,\partial _{t}\psi  &=&\left[ -{(1/2)}\left( \partial _{x}^{2}+\partial
_{y}^{2}+\partial _{z}^{2}\right) \right.   \nonumber \\
&&\left. +{(1/2)}\rho _{\perp }^{2}+V(z)+2\pi \,g|\psi |^{2}\right] \psi .
\label{GPE}
\end{eqnarray}%
Here $\psi (x,y,z,t)$ is the mean-field wave function, and $\rho _{\perp
}^{2}\equiv x^{2}+y^{2}$. Further, $g=2a_{s}N/a_{\bot }$is the nonlinearity
strength, with $a_{s}$ the \textit{s}-wave scattering length, $N$ the total
number of atoms in the condensate, and $a_{\bot }=\sqrt{\hbar /(m\omega
_{\bot })}$ the confinement radius imposed by the transverse harmonic
potential of frequency $\omega _{\bot }$, with $m$ the atomic mass. In Eq. (%
\ref{GPE}) length and time are measured in units of $a_{\bot }$ and $\omega
_{\bot }^{-1}$. As usual, $g>0$ and $g<0$ correspond to the repulsion and
attraction between atoms in the BEC, respectively, and $V(z)$ is a weak
axial potential, which may be present in addition to the strong transverse
confinement. Being interested in the stabilization mechanism that does not
require the extra potential, we set $V(z)=0$. Then, the 3D equation can be
reduced to the NPSE by means of \textit{ansatz} \cite{sala1}%
\begin{equation}
\psi (\rho _{\perp },z,t)={\frac{1}{\sqrt{\pi }\sigma (z,t)}}\exp {\left\{ -{%
\frac{r^{2}}{2\left[ \sigma (z,t)\right] ^{2}}}\right\} }\,f(z,t),
\label{ansatz}
\end{equation}%
where 1D wave function $f(z,t)$ is subject to the normalization condition, $%
\int_{-\infty }^{+\infty }\left\vert f(z,t)\right\vert ^{2}dz=1$. Following
Refs.\textbf{\ } \cite{sala1}-\cite{we}, one can eliminate the transverse
width, $\sigma ^{4}=1+{g|f|^{2}}$, arriving at the NPSE,
\begin{equation}
i\frac{\partial f}{\partial t}=\left[ -\frac{1}{2}\frac{\partial ^{2}}{%
\partial z^{2}}+V(z)+\frac{1+(3/2)g|f|^{2}}{\sqrt{1+g|f|^{2}}}\right] f.
\label{final}
\end{equation}%
In the case of $g=$ $\mathrm{const}$, stationary solutions are looked for as
$f=\exp \left( -i\mu t\right) \Phi (z)$, where $\mu $ is the chemical
potential, and real function $\Phi (z)$ obeys equation
\begin{equation}
\mu \ \Phi =\left[ -\frac{1}{2}\Phi ^{\prime \prime }+V(z)\Phi +{g}\frac{%
1+(3/2)\Phi ^{2}}{\sqrt{1+g\Phi ^{2}}}\right] \Phi .  \label{Phi}
\end{equation}%
Some numerical methods for simulations of the GPE and NPSE (with $g=$ const)
were presented in Ref. \cite{sala-numerics}.

In the case of the attractive nonlinearity, $g<0$, the form of Eq. (\ref%
{final}) implies that the amplitude of the wave function is limited from
above by a critical value,
\begin{equation}
\left\vert f\right\vert ^{2}<\left( \left\vert f\right\vert ^{2}\right) _{%
\mathrm{cr}}=1/|g|.  \label{dyn-coll}
\end{equation}%
A \textit{dynamical collapse} sets in, with transverse width $\sigma $
shrinking to zero and the solution developing a singularity in finite time,
as $\left\vert f\right\vert ^{2}$ approaches the critical value \cite%
{sala-soli}. In Ref. \cite{antipatici}, this was called \textit{%
two-dimensional primary collapse}, as it is related to the transverse 2D
dynamics.

In addition to the dynamical collapse, the NPSE also admits a \textit{static
collapse}, in the framework of stationary equation (\ref{Phi}): for $g<0$
and $V(z)=0$, this equation admits bright-soliton solutions only below the
critical value of the nonlinearity strength, $|g|\leq \left\vert
g_{c}\right\vert \equiv {4/3}$ \cite{sala-soli}. At $|g|\leq \left\vert
g_{c}\right\vert $, the axial density $|\Phi (z)|^{2}$ in the bright-soliton
solution is smaller than the critical value imposed by condition (\ref%
{dyn-coll}). In Ref. \cite{antipatici}, this kind of the collapse was called
\textit{three-dimensional primary collapse}, as it involves a
quasi-spherical 3D soliton. \ With regard to the definition of $g$, this
restriction determines the largest number of atoms possible in the soliton, $%
N<N_{\max }=(2a_{\perp }/3\left\vert a_{s}\right\vert )$. With $a_{\perp }$
and $\left\vert a_{s}\right\vert $ on the order of $\mathrm{\mu }$m and $0.1$
nm, respectively, which is typical for experiments in the $^{7}$Li
condensate \cite{soliton}, one has $N_{\max }\sim 10^{4}$ atoms.

The FRM technique which makes it possible to stabilize 2D matter-wave
solitons is based on the respective GPE, $i\,\partial _{t}\psi =\left[ -{%
(1/2)}\left( \partial _{x}^{2}+\partial _{y}^{2}\right) +2\pi \,g(t)|\psi
|^{2}\right] \psi ,$ where constant $g<0$ is replaced by a periodic
function, $g(t)=g_{\mathrm{dc}}+g_{\mathrm{ac}}\sin \left( \omega t\right) ,$%
with $\left\vert g_{\mathrm{dc}}\right\vert <g_{\mathrm{ac}}$, so
that $g(t)$ alternates between attraction and repulsion. The
stabilization requires the presence of the constant (``dc")
component which corresponds to the self-attraction on the average,
i.e., $g_{\mathrm{dc}}<0$.

The action of the FRM within the framework of the NPSE was not considered
before. To explore this situation, we take $g(t)$ as indicated above,
arriving at the following modification of Eq. (\ref{final}):
\begin{equation}
i\frac{\partial f}{\partial t}=\left[ -\frac{1}{2}\frac{\partial ^{2}}{%
\partial z^{2}}+\frac{1+(3/2)\left[ g_{\mathrm{dc}}+g_{\mathrm{ac}}\sin
\left( \omega t\right) \right] |f|^{2}}{\sqrt{1+\left[ g_{\mathrm{dc}}+g_{%
\mathrm{ac}}\sin \left( \omega t\right) \right] |f|^{2}}}\right] f.
\label{1DFeshbach}
\end{equation}%
Our objective is to identify a region in parameter space $\left\{ g_{\mathrm{%
ac}},g_{\mathrm{dc}},\omega \right\} $ where the solitons subjected to the
``management" represent stable solutions of Eq. (\ref%
{1DFeshbach}).

\section{Results}

Localized solutions to Eq. (\ref{1DFeshbach}) were categorized as stable
ones if, in direct simulations, they featured persistent pulsations,
avoiding collapse or decay up to $t=5000$ (in some cases, the stability was
checked up to $t=80000$). However, the application of this criterion to the
case of $\omega \lesssim 0.5$ is complicated by the fact that, under the
action of the low-frequency management, the soliton tends to develop an
apparently chaotic behavior, although without a trend to decay, see below.
Fixing the time interval as that comprising a large number of periods,
simulations become increasingly more difficult for $\omega \rightarrow 0$.

The simulations we performed by means of the Crank-Nicolson algorithm with
open-ended boundary conditions. The initial state was $f_{0}(z)=\mathrm{sech}%
(z)$, that was refined by the integration of Eq.~(\ref{1DFeshbach}) with $g_{%
\mathrm{ac}}=0$ in imaginary time \cite{sala-numerics}. The so generated
configuration was then used as the input to simulate Eq.~(\ref{1DFeshbach})
in real time.

Typical examples of stable and unstable solutions are shown in Figs. \ref%
{fig:1} and \ref{fig:2} (in the model with $g_{\mathrm{ac}}=0$, the
respective soliton is stable). In the latter case, a gradually growing
amplitude of the soliton attains critical value (\ref{dyn-coll}) at finite $t
$, which implies the onset of the dynamical collapse. It happens when the
argument of the square root in Eq. (\ref{1DFeshbach}) becomes zero, i.e.,
the transverse width of the soliton shrinks to zero. Note that, with $g=g(t)$%
, critical density (\ref{dyn-coll}) of the axial wave function does not
necessarily correspond to the maximum of $g(t)$. Indeed, in Fig. \ref{fig:2}%
(a) the collapse happens at $g(t)$ smaller than its maximum value, $|g_{%
\mathrm{dc}}|+|g_{\mathrm{ac}}|$.

\begin{figure}[tbp]
\includegraphics[width=3cm,clip]{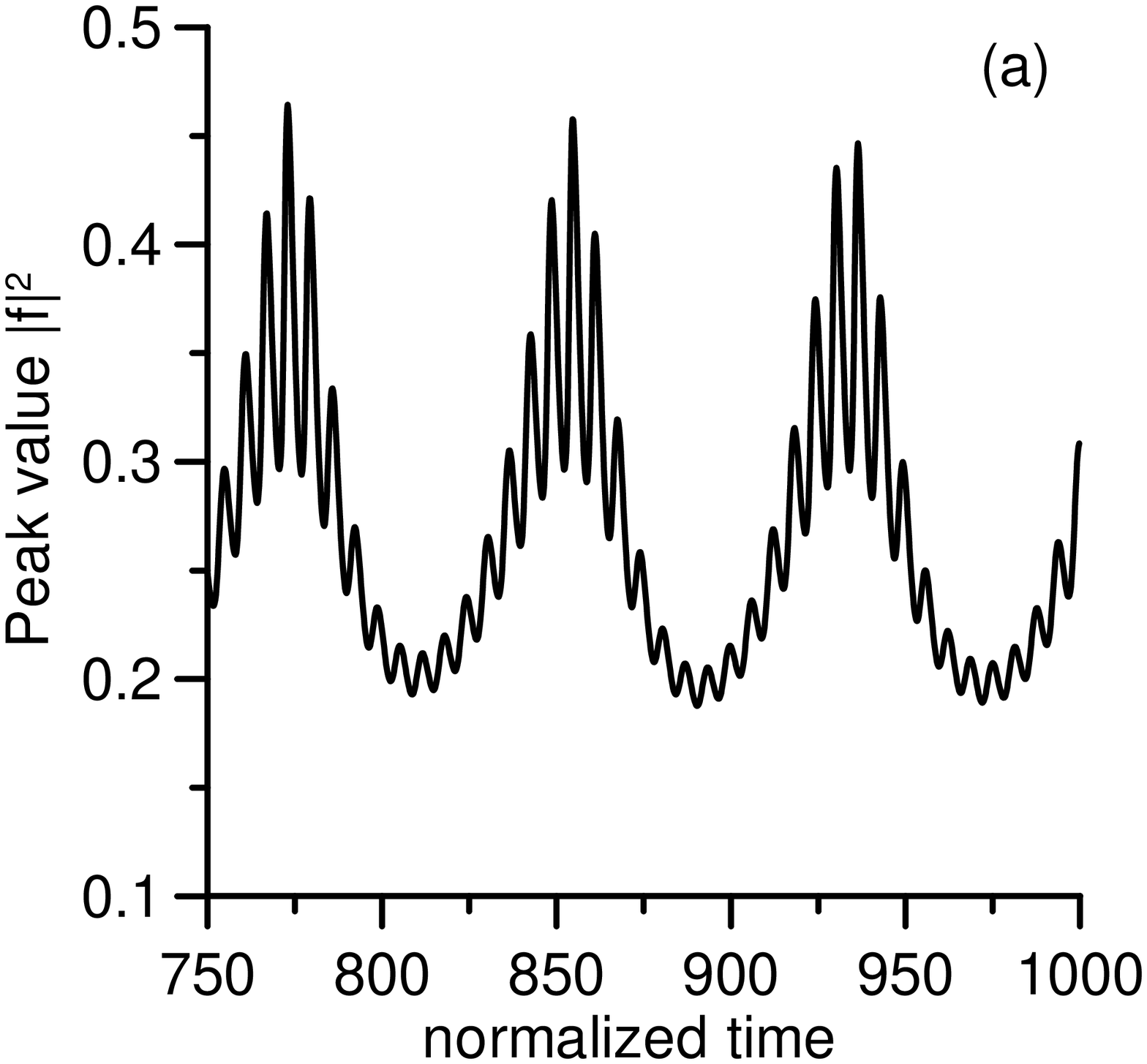}%
\includegraphics[width=3cm,clip]{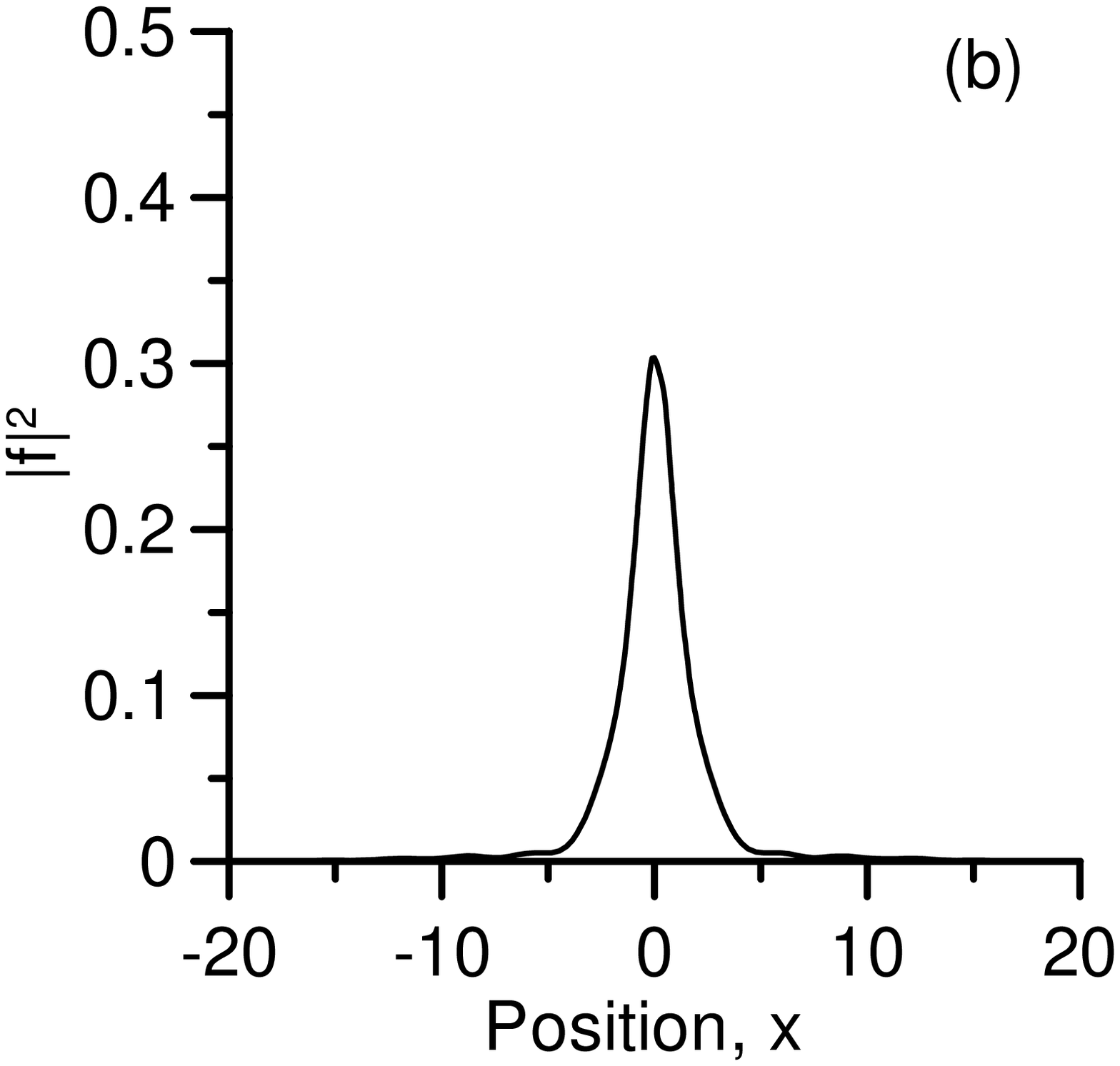}
\caption{A typical example of a stable soliton solution to Eq.~(\protect\ref%
{1DFeshbach}). (a) The evolution of the soliton's amplitude in time. (b) A
snapshot of the soliton at $t=100$. The integration step is $\Delta t=0.05$,
and the size of the integration domain is $L=200$. Parameters are $g_{%
\mathrm{dc}}=-1$, $g_{\mathrm{ac}}=1.5$, $\protect\omega =1$. }
\label{fig:1}
\end{figure}

\begin{figure}[tbp]
\centering
\includegraphics[width=4.5cm,clip]{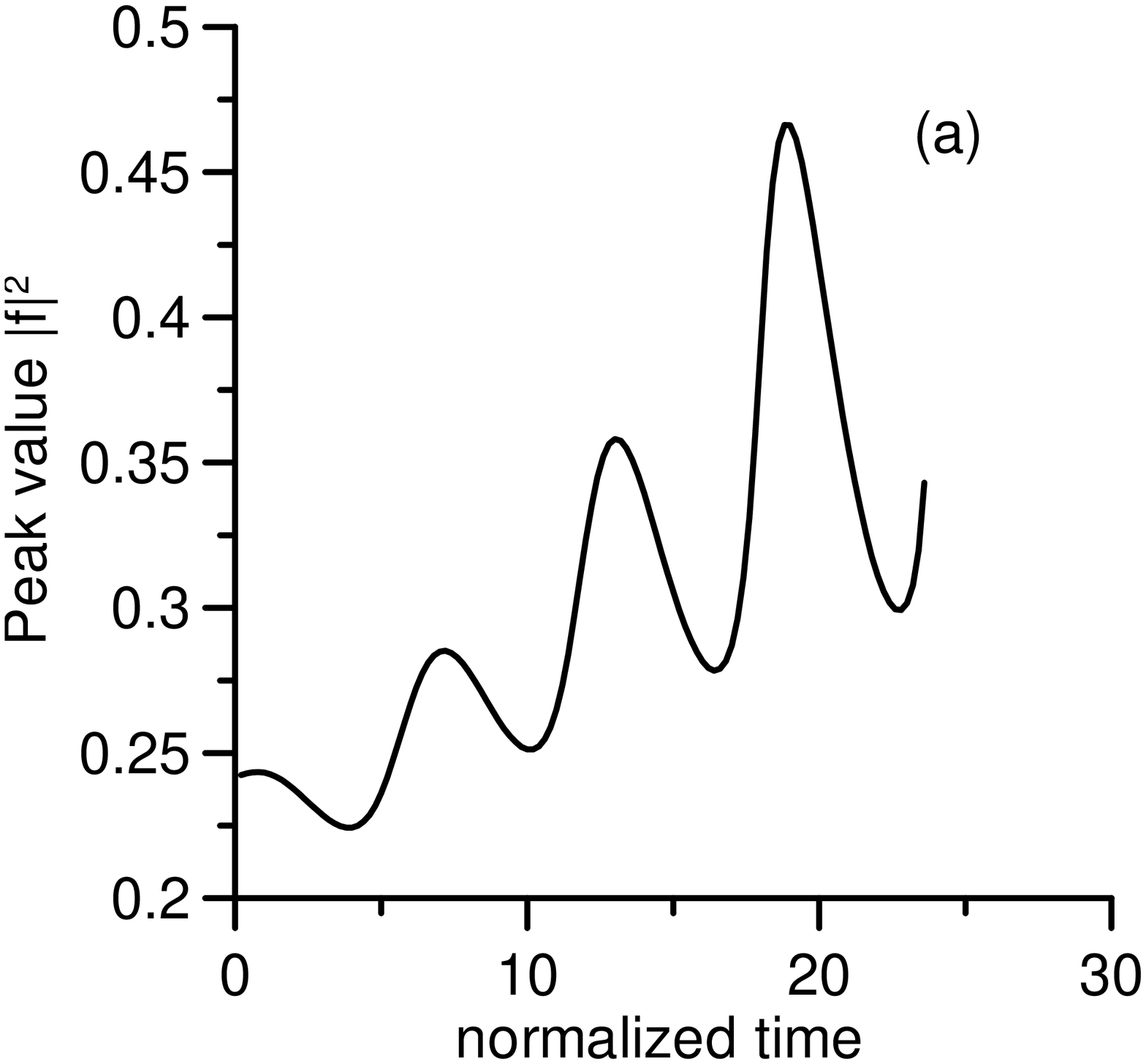} %
\includegraphics[width=3.5cm,clip]{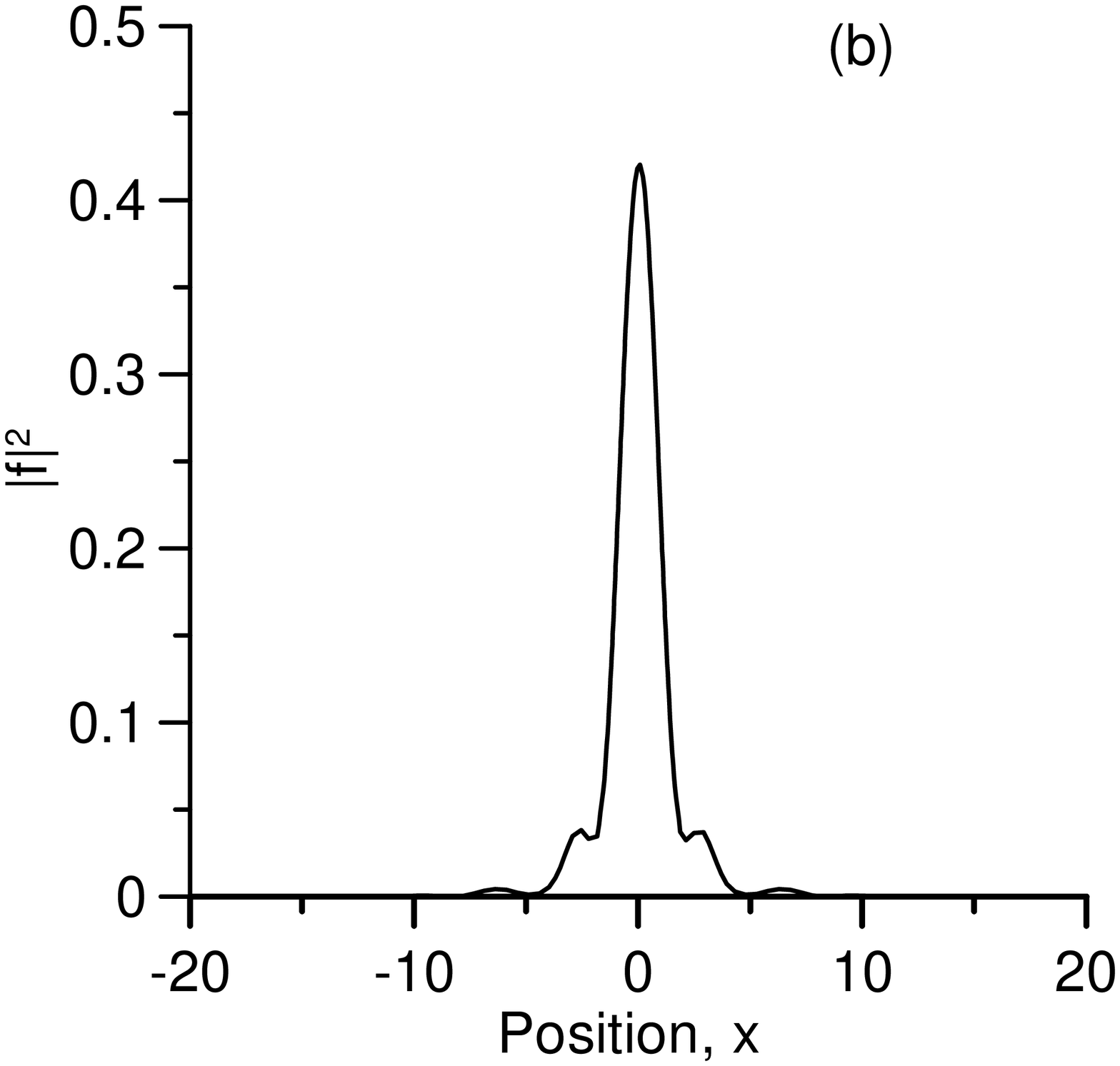}
\caption{A typical example of the collapsing soliton. (a) The evolution of
the amplitude up to $t\approx 25$, when it reaches critical value (\protect
\ref{dyn-coll}). (b) A snapshot of the soliton just before the onset of the
collapse. Parameters are the same as in Fig. \protect\ref{fig:1}, except for
$g_{\mathrm{ac}}=2$.}
\label{fig:2}
\end{figure}

\begin{figure}[tbp]
\centering
\includegraphics[width=7cm,clip]{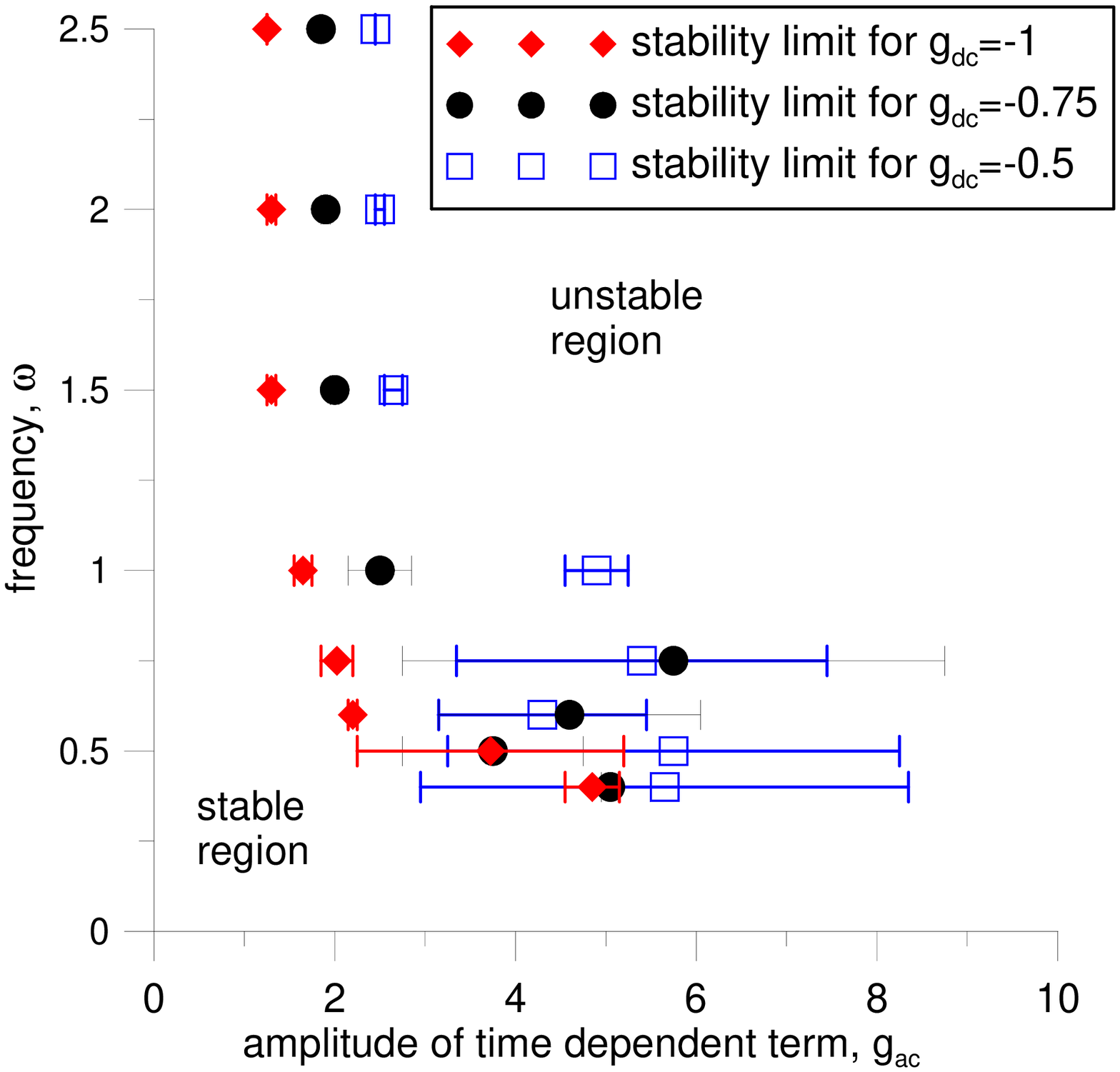}
\caption{(Color online). Stability borders in the plane of the
time-modulation parameters, $\left( \protect\omega ,g_{\mathrm{ac}}\right) $%
, as obtained from systematic simulations for different fixed values of $g_{%
\mathrm{dc}}$. In cases when the threshold depends upon the integration time
(see Fig. \protect\ref{fig:4}), the respective symbol corresponds to the
mean value, with the error bars given as per the respective semi-dispersion.}
\label{fig:3}
\end{figure}

Results of systematic simulations are summarized in the form of the
stability diagram displayed in Fig. \ref{fig:3}. The stability thresholds
shown in the figure, i.e. the maximum value of $g_{\mathrm{ac}}$ admitting
stable solitons, were found by slowly increasing $g_{\mathrm{ac}}$ in steps
of $\Delta g_{\mathrm{ac}}=0.1$, until the instability was attained. The
shape of the stability domain in the plane of the management parameters, $%
\left( g_{\mathrm{ac}},\omega \right) $, is roughly similar to that which
was found in management models of a different type, with the time-periodic
modulation applied not to the nonlinearity, but to the optical-lattice
potential, which is necessary for the existence of stable solitons in those
cases. These include the 1D model for gap solitons, with a positive
scattering length \cite{Thawatchai1}, and the 2D GPE with a negative
scattering length and 1D or 2D periodic potential, that stabilizes TSs in
the respective settings \cite{Thawatchai2}. As in those works, one may
expect that here, at very large values of $\omega $, the stability region
will start to expand in the direction of larger values of $g_{\mathrm{ac}}$,
as in the limit of $\omega \rightarrow \infty $ the ac term averages to zero.

The stability borders in Fig. \ref{fig:3} are not extended to $\omega <0.5$,
as in the region of the low-frequency modulation the solitons feature
persistent but apparently \emph{chaotic} evolution. In fact, the stability
domain is well defined for $\omega \geq 1$, while in the intermediate
region, $0.5\leq \omega <1$, the randomness of the soliton evolution makes
the stability border dependent on the integration time -- see Fig. \ref%
{fig:4}, which demonstrates a natural trend to a \textit{decrease} of the
effective instability threshold with the increase of the evolution time, if
the threshold is sensitive to it at all.

\begin{figure}[tbp]
\centering
\includegraphics[width=5.5cm,clip]{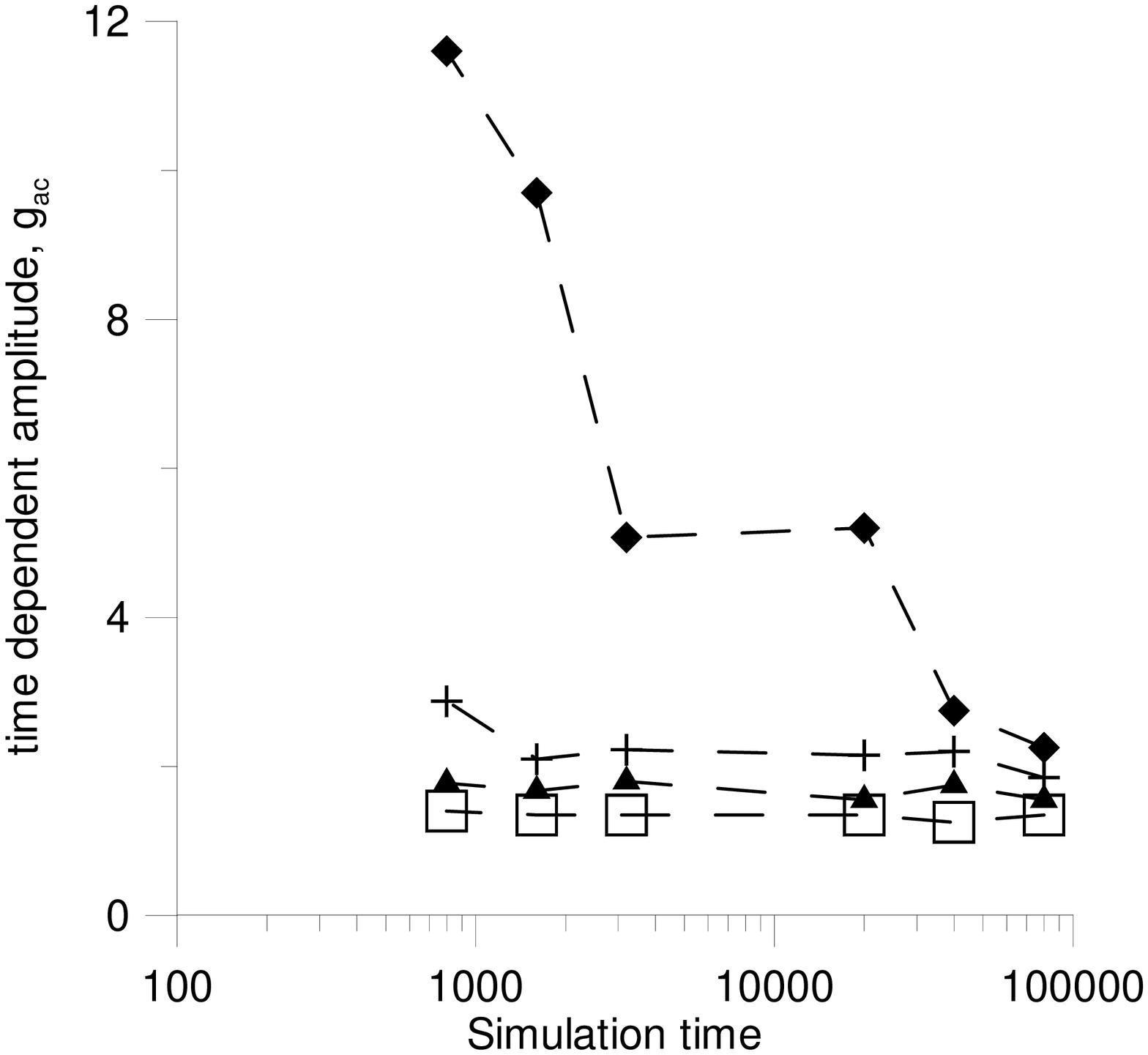}
\caption{Dependence of $g_{\mathrm{ac}}$ at the soliton's instability border
on the integration time, for different modulation frequencies and $g_{%
\mathrm{dc}}=-1$. Symbols denote the stability limits at different
frequencies: $\protect\omega =0.5$ (diamonds), $=0.75$ (crosses), $1$
(triangles), $1.5$ (squares).}
\label{fig:4}
\end{figure}

\section{Conclusion}

We have used the NPSE, i.e., the 1D mean-field equation for tightly trapped
BEC, with the nonpolynomial nonlinearity admitting the onset of the collapse
in the framework of the 1D description, for the study of the stabilization
of solitons by means of the FRM\ (Feshbach-resonant-management) technique.
The results were reported in the form of stability diagrams in the plane of
the management parameters, $\left( g_{\mathrm{ac}},\omega \right) $. The
stability domain is roughly similar to that reported in \textit{%
linear-management} models for 1D gap solitons and 2D TSs (Townes solitons),
supported by optical lattices subjected to the time-periodic modulation.
However, stability domains of such a form have not been reported before in
models of the \textit{nonlinearity management}. At small values of the
modulation frequency, the stability border becomes fuzzy, as solitons
feature chaotic evolution in that case.

\end{document}